# Making Atomic-Level Magnetism Tunable with Light at Room Temperature

V.O. Jimenez[1,*], Y.T.H. Pham[1], D. Zhou[2], M.Z. Liu[2], F.A. Nugera[1], V. Kalappattil[1], T. Eggers[1], K. Hoang[3], D.L. Duong[4], M. Terrones[2], H.R. Gutiérrez[1], and M.H. Phan[1,*]

[1] Department of Physics, University of South Florida, Tampa, FL 33620, USA

[2] Department of Physics, The Pennsylvania State University, University Park, PA 16802, USA

[3] Center for Computationally Assisted Science and Technology and Department of Physics, North Dakota State University, Fargo, ND 58108, USA

[4] Department of Physics, Montana State University, Bozeman, MT 59717, USA

**The capacity to manipulate magnetization in two-dimensional dilute magnetic semiconductors (2D-DMSs) using light, specifically in magnetically doped transition metal dichalcogenide (TMD) monolayers (*M*-doped $TX_2$, where $M$ = V, Fe, Cr; $T$ = W, Mo; $X$ = S, Se, Te), may lead to innovative applications in spintronics, spin-caloritronics, valleytronics, and quantum computation. This Perspective paper explores the mediation of magnetization by light under ambient conditions in 2D-TMD DMSs and heterostructures. By combining magneto-LC resonance (MLCR) experiments with density functional theory (DFT) calculations, we show that the magnetization can be enhanced using light in V-doped TMD monolayers (e.g., V-$WS_2$, V-$WSe_2$, V-$MoS_2$). This phenomenon is attributed to excess holes in the conduction and valence bands, as well as carriers trapped in magnetic doping states, which together mediate the magnetization of the semiconducting layer. In 2D-TMD heterostructures such as $VSe_2/WS_2$ and $VSe_2/MoS_2$, we demonstrate the significance of proximity, charge-transfer, and confinement effects in amplifying light-mediated magnetism. This effect is attributed to photon absorption at the TMD layer (e.g., $WS_2$, $MoS_2$) that generates electron-hole pairs mediating the magnetization of the heterostructure. These findings will encourage further research in the field**

**of 2D magnetism and establish a novel direction for designing 2D-TMDs and heterostructures with optically tunable magnetic functionalities, paving the way for next-generation magneto-optic nanodevices.**

***Corresponding authors:** valeryortizj@usf.edu (V.O.J); phanm@usf.edu (M.H.P)

## 1. Introduction

Magnetic semiconductors present an exceptional platform for the development of a new generation of highly efficient spintronic devices, including spin field-effect (SFE) transistors [1-7]. In contrast to conventional field-effect transistors that rely on electron charge, SFE transistors utilize electron spin and its alignment (up or down) within a magnetic semiconductor to encode binary information, facilitating rapid information transmission with minimal power consumption [1,4]. To diminish the size of such devices, it is crucial to reduce the dimensions of semiconductor materials [4,7-9]. As dimensions decrease, novel physical properties emerge, and new potential applications arise. However, miniaturization to the nanoscale might substantially impair their performance due to current leakage, rendering such devices inapplicable in ultrafast electronic nanodevices, particularly in supercomputers or future quantum computers [1,7]. For most magnetic semiconductors, ferromagnetic properties are largely weakened or even lost when their thickness is reduced to the atomic level or two-dimensional (2D) limit [1,2,7].

Recent advances in the domain of 2D van der Waals (vdW) magnetic materials have yielded unprecedented opportunities for exploiting atomically thin magnets [10-14] and heterostructures [15-27] with tunable magnetic, magnetoelectric, and magneto-optic properties. Among the identified 2D vdW magnets, the atomically thin intrinsic magnetic semiconductors $CrI_3$ and $Cr_2Ge_2Te_6$ have been extensively studied for their novel magnetoelectric, magneto-optic, and spin transport properties [10,11,28-37]. Their magnetic functionalities can also be modulated by external stimuli (electric gating, strain, light) [38-47]. Regrettably, these 2D semiconductors exhibit magnetic ordering at low temperatures (< 50 K), limiting their practical implementation. Consequently, there is a growing demand for the development of 2D magnetic semiconductors that exhibit ferromagnetic ordering at ambient temperatures, under which most electronic devices operate.

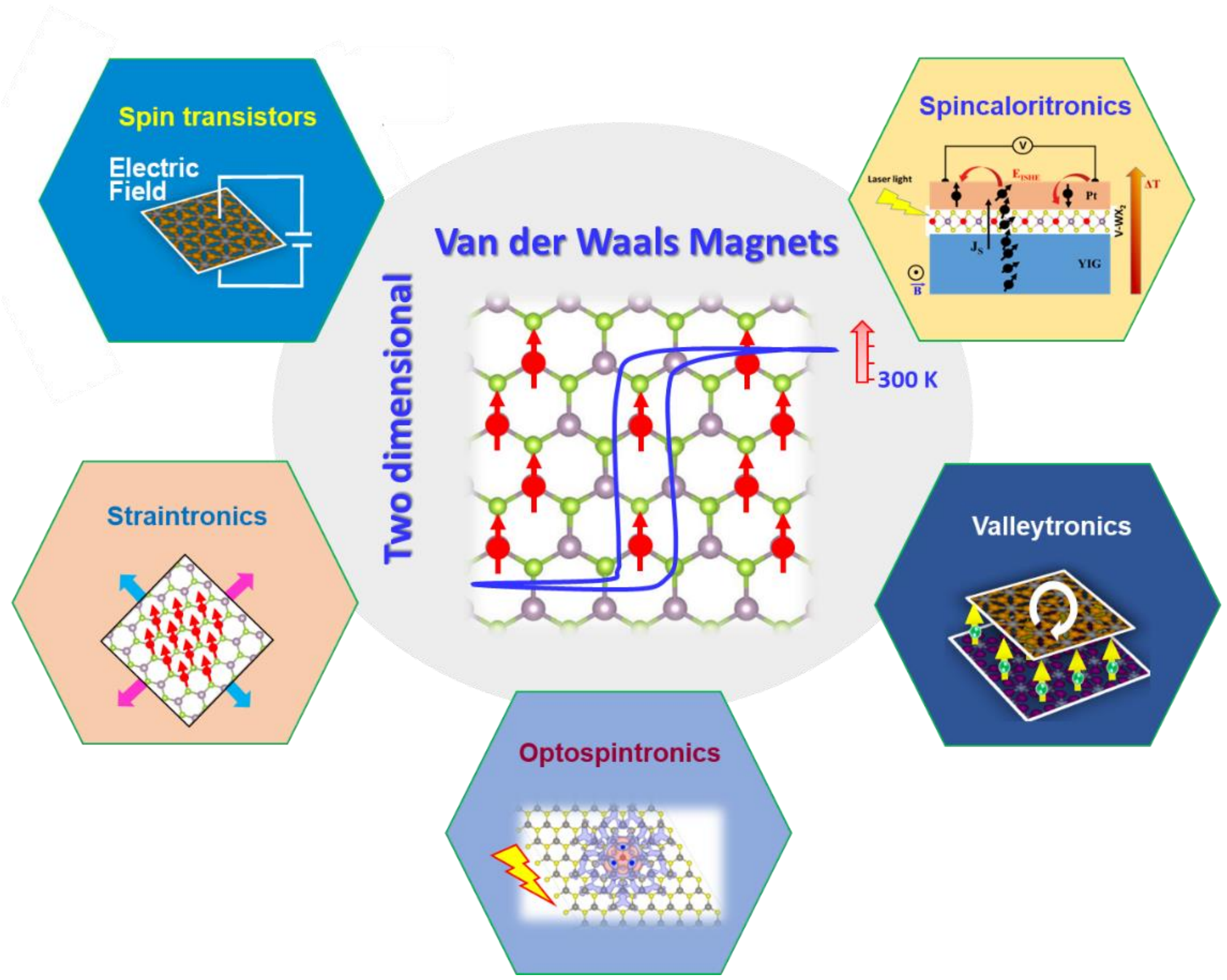

**Figure 1.** Perspectives of novel device applications of atomically thin magnetic transition metal dichalcogenides and their heterostructures.

Two-dimensional transition metal dichalcogenides (2D-TMDs) $TX_2$ ($T$ = W, Mo; $X$ = S, Se, Te) are central to numerous vital device applications such as field-effect transistors, photodetectors, photon emitters, valleytronics, and quantum computers [3,4,7,8,48,49]. Despite the presence of certain 2D-TMDs such as $VSe_2$ [13], $MnSe_2$ [14], and $CrSe_2$ [50], which exhibit ferromagnetic ordering near room temperature but are *metallic*, the majority of *semiconducting* 2D-TMDs including $WSe_2$, $WS_2$, and $MoS_2$ monolayers are non-magnetic or diamagnetic in nature [51]. Recent studies have shown that introducing small quantities of magnetic transition metal atoms (e.g., V, Fe, Co, Cr, Mn) can induce long-range ferromagnetic order in these 2D-TMDs at room temperature [52-73]. This approach presents a promising strategy to integrate extrinsic magnetic properties into atomically thin TMD semiconductors, giving rise to a novel class of two-dimensional dilute magnetic semiconductors (2D-DMSs). Due to their high-quality interfaces and weakly coupled interlayer interactions, 2D-TMDs with desirable properties can be easily stacked together, creating 2D vdW heterostructures with unique properties otherwise absent in their individual components [7,77-86]. Their potential for next generation spintronic, opto-spintronic, opto-spin-caloritronic, and valleytronic device applications has been emphasized, owing to their atomically thin nature and integrated opto-electro-magnetic properties [7,83-86]. Figure 1 illustrates potential applications of 2D-TMD DMSs and their heterostructures.

The true appeal of 2D-TMD DMSs for these applications stems from their magnetic tunability in response to external stimuli (electric gating, light, strain). Magnetic state tunability of a 2D-TMD can range from enhancing its magnetic moment, tuning its Curie temperature to inducing magnetism in non-magnetic materials through chemical doping [52,53], defect engineering [60], phase change or structure engineering [75,87,88], interface engineering [77-82], or applying external stimuli [56,57,89]. Among these approaches, the capacity to modulate the magnetic moments of 2D-TMDs reversibly, using external stimuli such as electric gating or light, appears to accommodate the ever-increasing demands for multifunctional sensing devices, information storage, and quantum computing technologies. Our discovery of tunable room temperature ferromagnetism in atomically thin V-doped TMD (V-$WS_2$, V-$WSe_2$, V-$MoS_2$) semiconductors [52,53] has provided a new possibility for controlling their magnetic and magneto-electronic properties through optical means [56,90]. By combining the light-tunable magnetism of 2D-DMSs [56,90] and the spin Seebeck effect [91], we have proposed an innovative strategy for the optic control of thermally driven spin currents across magnet/metal interfaces in spincaloritronic devices, potentially establishing a new subfield dubbed “Opto-Spin-Caloritronics” [84]. To fully exploit the optically tunable magnetic properties of 2D-TMD DMSs and their heterostructures for spintronics, spin-caloritronics, straintronics, and valleytronics (Fig. 1), it is essential to comprehend the underlying mechanisms of light-mediated magnetism in these 2D systems.

In this Perspective article, we demonstrate how light modulates the magnetization in 2D-TMD DMSs (V-$WS_2$, V-$WSe_2$, V-$MoS_2$) and associated heterostructures ($VSe_2/MoS_2$, $VSe_2/WS_2$), using an ultrasensitive magneto-LC resonance (MLCR) magnetometer. Supplemented by DFT calculations, these findings enable us to propose innovative design strategies for novel 2D-TMD DMSs and heterostructures with enhanced light-tunable magnetic functionalities suitable for modern nanodevice applications. Figure 2 presents two promising approaches for creating such 2D-TMDs and heterostructures, with their optically tunable magnetic properties discussed herein.

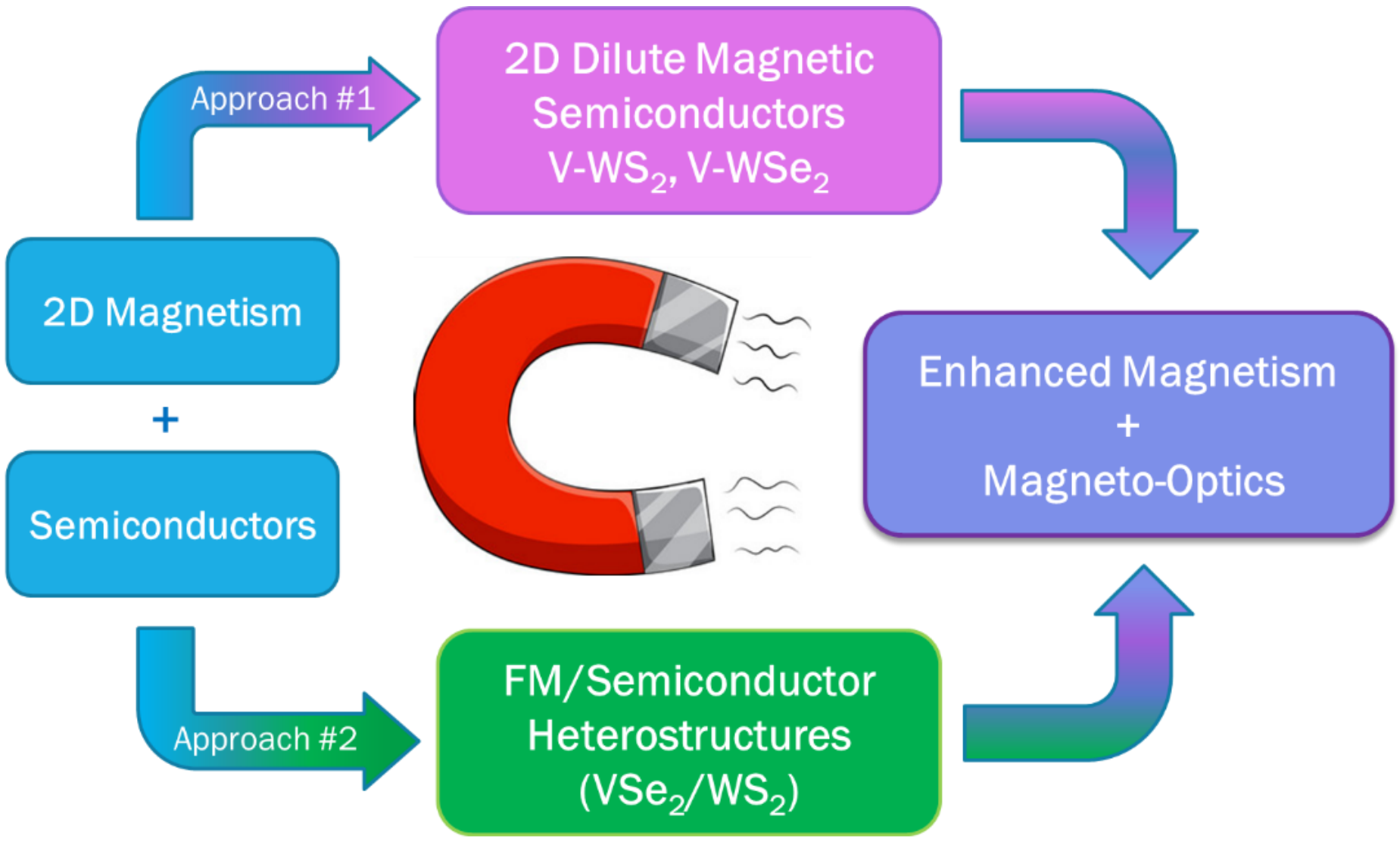

**Figure 2. Approach 1:** Introducing small amounts of magnetic atoms (e.g., V, Fe, Cr) into semiconducting TMD monolayers (e.g., $WS_2$, $WSe_2$, $MoS_2$) creates a novel class of 2D dilute magnetic semiconductors (e.g., V-$WS_2$, V-$WSe_2$, V-$MoS_2$). **Approach 2**: Interfacing 2D-TMD magnets (e.g., $VSe_2$, $VS_2$, $MnSe_2$, $CrSe_2$) with 2D-TMD semiconductors (e.g., $WS_2$, $WSe_2$, $MoS_2$) create a novel class of magneto-optic 2D van der Waals heterostructures (e.g., $VSe_2/WS_2$, $VS_2/MoS_2$, $MnSe_2/WSe_2$, $CrSe_2/WSe_2$).

The paper is structured as follows: we first introduce the MLCR magnetometry technique and then use this technique to demonstrate the light-mediated magnetism effects in the 2D-TMD DMSs and related heterostructures. We will also discuss emerging opportunities and challenges in the field of study.

## 2. Magneto-LC Resonance Magnetometry for Probing Light-Mediated Magnetism

As ferromagnetic signals become exceedingly weak in atomically thinned magnetic systems, probing small changes in magnetization of the material subject to external stimuli, such as light, presents a considerable challenge [39,41,43,92-103]. Superconducting quantum interference devices (SQUID) can measure the magnetization of 2D materials [52,53], but they are not well-suited for real-time measurements while simultaneously illuminating the samples with light [93]. Transport measurements on 2D materials also pose difficulties, as the size of the electrical contacts is often larger compared to the surface area of the sample [7]. Optical methods based on the magneto-optic Kerr effect (MOKE), time-resolved Faraday rotation, and reflectance magneto-circular dichroism (RMCD) have been successfully employed to characterize the light-mediated magnetic properties of 2D materials such as $Fe_3GeTe_2$ [92], $Cr_2Ge_2Te_6$ [40], and $CrI_3$ [39]. However, the use of high laser powers in these methods may cause local heating and consequently thermal instability – a significant source of noise. The limitations of these techniques necessitate the development of a new approach to measure 2D magnetization in real time as external stimuli, such as light, are applied.

To probe the light-induced magnetization of an atomically thin magnetic film, we have developed a novel magneto-LC resonance (MLCR) magnetometer with ultrahigh magnetic field sensitivity (pT regime) [56,90].

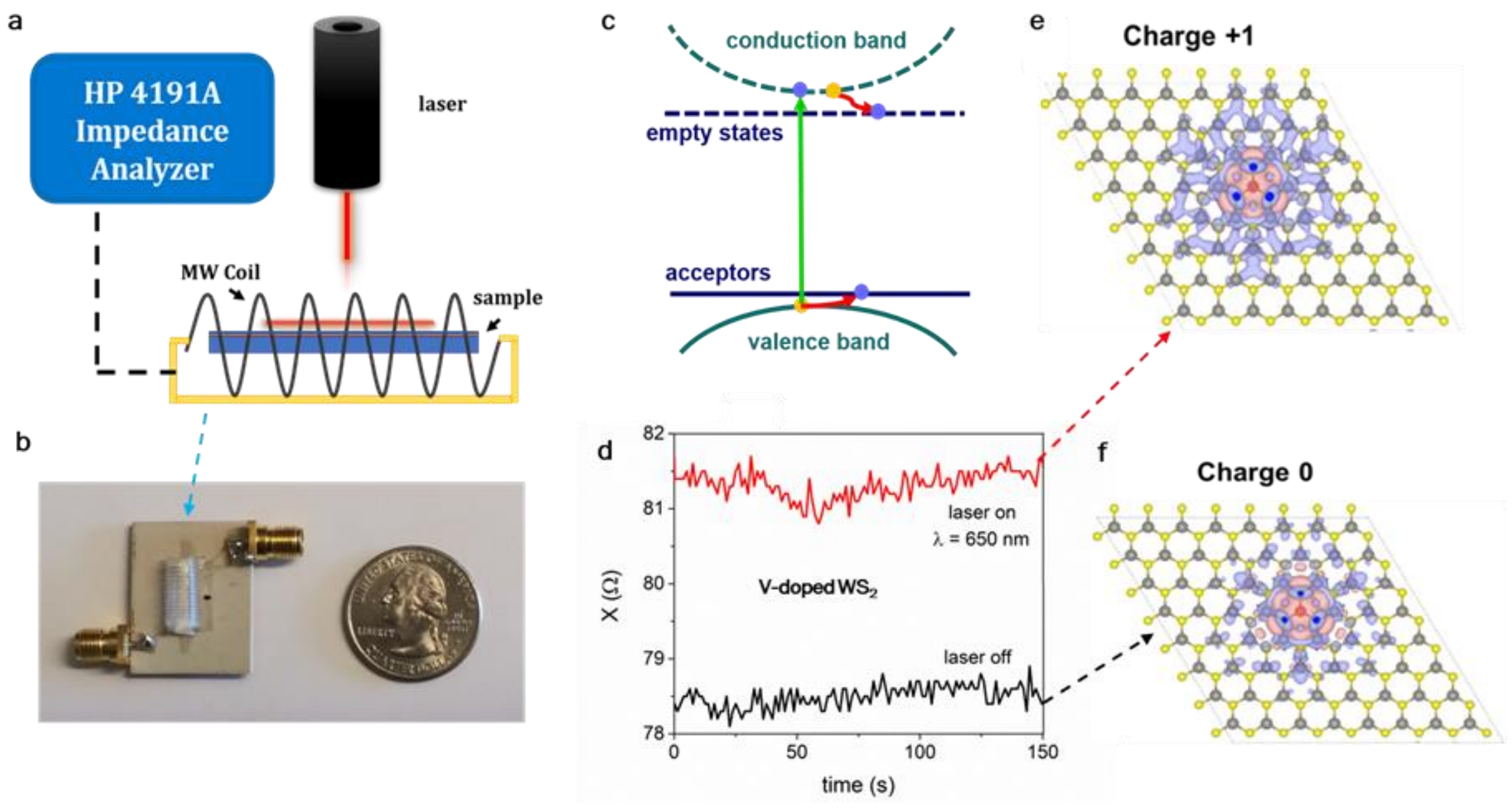

**Figure 3.** (**a**) Schematic of the MLCR measurement setup. Changes in magnetic reactance corresponding to the magnetic permeability or magnetization of a magnetic sample due to light irradiation are monitored in real time by a soft magnetic coil. The coil is mounted on a test fixture composed of a copper clad dielectric material, with its two ends soldered into the inner pin of SMA ports, as depicted in (**b**). (**c**) Schematic of the photon-induced magnetism effect when light illuminates a 2D-TMD; (**d**) Reactance as a function of time when the laser is off and on; The projected magnetic moment along the c-axis of the V-$WS_2$ monolayer upon a single hole injection (**e**) relative to no charge injection (**f**). Both MLCR experiments (**d**) and DFT calculations (**e**,**f**) confirm that the magnetization of the V-$WS_2$ monolayer increases upon light illumination.

Figure 3a presents a schematic of the light-mediated magnetization measurement system using the principle of MLCR [104,105]. The MLCR design draws inspiration from conventional magneto-inductive coils and the sensitivity of the giant magneto-impedance (GMI) effect [106], which has proven useful for detecting ultra-small magnetic fields in biosensing applications [107] and for structural health monitoring [108]. The sensor is constructed from a Co-rich soft magnetic microwire exhibiting very high GMI ratios and magnetic field sensitivity. The melt-extracted amorphous microwire, with a nominal composition of $Co_{69.25}Fe_{4.25}Si_{13}B_{12.5}Nb_1$ and diameter of approximately 60 μm, is wound into a 15-turn, 10-mm-long coil with a 5 mm internal diameter. The coil is mounted on a test fixture made of a copper clad dielectric material (Fig. 3b). The two ends of the coil are soldered onto the inner pin of SMA ports, which are connected to a coaxial cable and terminated with a 50-Ω cap. The coil is then driven by a frequency in the MHz range (~118 MHz, near the coil's LC resonance), and the impedance ($Z$), resistance ($R$), and reactance ($X$) are measured using an HP 4191A impedance analyzer.

The operating principle of the magnetic microwire coil (MMC) can be described using lumped element circuit theory. A simple model for a coil sensor is a lumped element representation of a non-ideal inductor. Winding cylindrical conductors close to each other introduces parasitic elements $R_{par}$ and $C_{par}$, such that the non-ideal inductor can be represented as a series combination of an ideal inductor $L$ and $R_{par}$, in parallel with $C_{par}$. The impedance of the coil $Z_{coil}$ can then be written as

$$Z_{coil} = Z_{R_{par}} + Z_{C_{par}} \tag{1}$$

$$Z_{coil} = \frac{1}{\frac{1}{R_{par}+j\omega L}+\frac{1}{-j/\omega C_{par}}} \tag{2}$$

$$Z_{coil} = \frac{R_{par}+j\omega\left[L\left(1-\omega^2 LC_{par}\right)-C_{par}R_{par}^2\right]}{\left(1-\omega^2 LC_{par}\right)^2+\left(\omega C_{par}R_{par}\right)^2} \tag{3}$$

where $Z_{coil}$ is the impedance of the coil, $\omega$ is the angular frequency, and $j$ is the imaginary unit. Resonance occurs when the inductive reactance ($X_L$) and the capacitive reactance ($X_C$) have equal magnitudes but differ in phase by 180 degrees. In this case, minimal current flows through the wire, the impedance of the coil becomes very large, and self-resonance is achieved. This resonance frequency is given by:

$$f_0 = \frac{\sqrt{1-(R_{par}^2 C_{par}/L)}}{2\pi\sqrt{LC_{par}}} \tag{4}$$

The reactance of the coil is of particular interest to utilize the coil for detecting changes in the magnetic permeability within its core. The impedance of the coil has the general form:

$$Z_{coil} = R + jX. \tag{5}$$

Therefore, we can extract the reactance from the imaginary component of Eq. (3):

$$X_{\text{coil}} = \frac{\omega[\mathrm{L}(1-\omega^2 \mathrm{L}C_{par}) - C_{\text{par}} R_{par}^2]}{(1-\omega^2 \mathrm{L}C_{par})^2 + (\omega C_{par} R_{par})^2}. \tag{6}$$

To measure the magnetic permeability or magnetization of a magnetic thin film, such as a 2D-TMD DMS, subjected to an external stimulus such as light, the magnetic film is positioned within the core of the coil, and the reactance of the coil is measured during light irradiation. According to Eq. (6), the reactance ($X$) of the coil is strongly dependent on the induction ($L$) through the coil's core. Since the film is ferromagnetic, it will alter the relative permeability of the space within the coil, thereby changing the magnetic flux through the coil and consequently the reactance of the coil. As the microwire itself is ferromagnetic, the magnetization of the film will also lead to a change in the effective permeability of the microwire. Thus, the reactance of the sensor depends on this effective permeability, $X = X(\mu_{eff})$. Changes in the permeability of the film upon light illumination (i.e., the presence of additional holes mediating the magnetism in 2D-TMDs) will influence the effective permeability of the coil, which can be accessed through the change in its reactance: $\Delta X = X(\mu_{eff}, laser\ on) - X(\mu_{eff}, laser\ off)$. This change in reactance ($\Delta X$) is proportional to the change in magnetization $\Delta M$ of the film upon light illumination, as illustrated in Fig. 3c-f for the case of a V-doped $WS_2$ monolayer. Using the MLCR method, we investigate the optically tunable magnetic properties of selected 2D-TMD DMSs and their heterostructures, with some of the results presented and discussed below.

## 3. Light-Tunable Two-Dimensional Magnetism

### *3.1. Magnetically Doped Transition Metal Dichalcogenide Monolayers*

Among 2D vdW materials, 2D-TMDs are a fertile ground for novel quantum phenomena including nontrivial electronic topology, non-saturating giant magnetoresistance, and topological field-effect transistors [1,3,4,7]. Recently, TMD monolayers (e.g., $WS_2$, $WSe_2$, $MoS_2$) doped with magnetic transition metal atoms (e.g., V, Fe, Co, Cr) have been reported to exhibit room-temperature ferromagnetic order, emerging as a novel class of 2D-DMSs [52-71]. We have discovered tunable room-temperature ferromagnetism in V-$WS_2$ and V-$WSe_2$ monolayers by varying the V-doping concentration [52,53]. Figure 4 highlights some notable features of these 2D magnets.

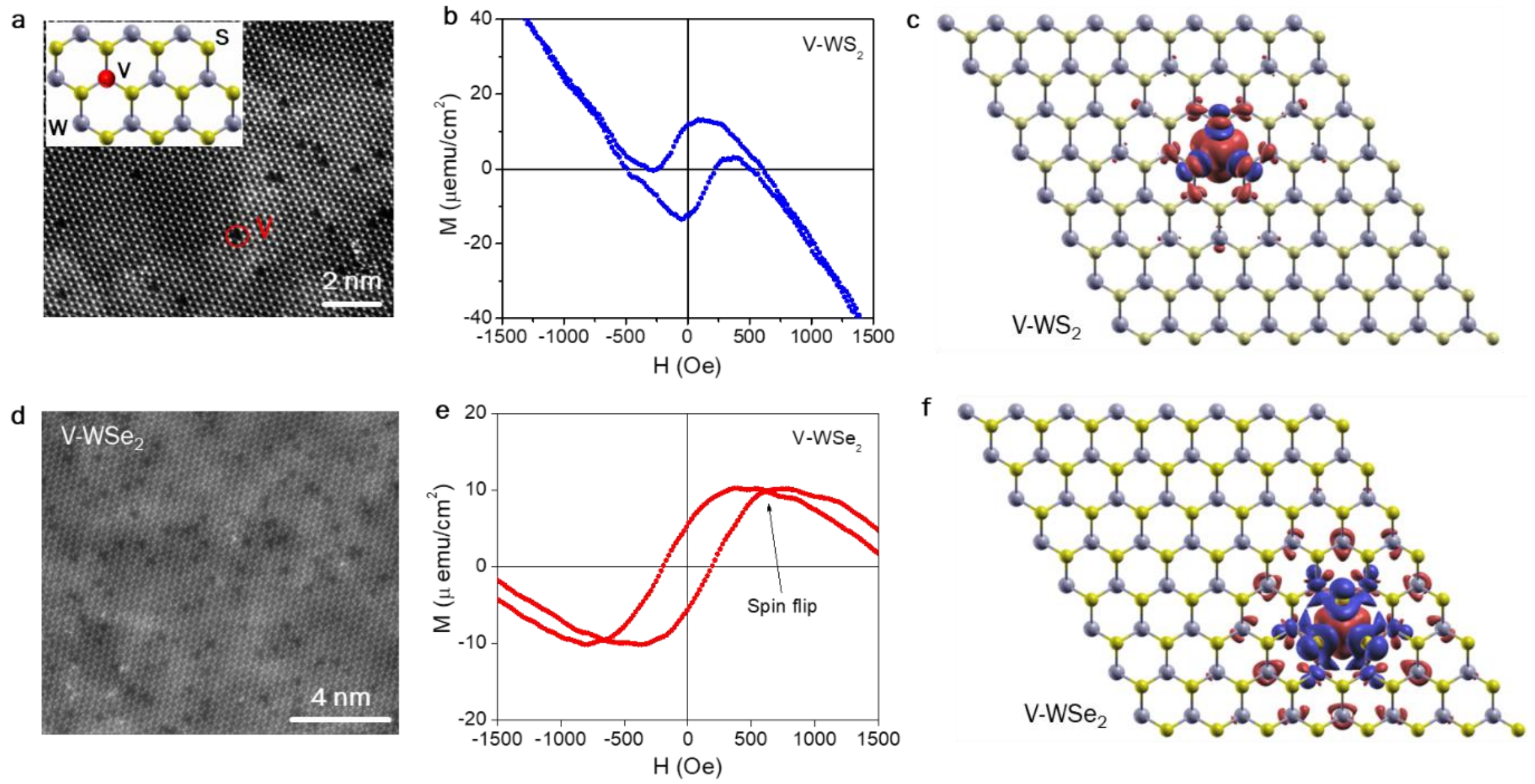

**Figure 4.** HRTEM images, magnetic hysteresis loops *M*(*H*), and spin configurations of (**a,b,c**) 2 at.% V-doped $WS_2$ and (**d,e,f**) 4 at.% V-doped $WSe_2$ monolayers, respectively. Within the V-$WS_2$ monolayer, the magnetic moment of V (replacing one W) is *ferromagnetically* coupled to the magnetic moments induced at the nearest and further distance W sites (all red, see panel **c**). In the V-$WSe_2$ monolayer, the magnetic moment of V is *antiferromagnetically* coupled to the magnetic moments at the nearest distance W sites (blue vs. red, see panel **f**) but *ferromagnetically* coupled to those at the further distance W sites (red, see panel **f**). It is the antiferromagnetic (AFM) coupling between the V and W spins at the nearest distance that leads to the thermally induced spin flipping phenomenon manifested as a crossover of magnetization in the $M(H)$ loop. Panels (a,b) are taken with permission from Ref. [52]; Panels (d,e) are taken with permission from Ref. [53].

By varying V-doping concentrations, we have demonstrated an enhanced magnetization and achieved the highest doping levels ever attained for atomically thin vanadium doped TMDs (~2 at.% for V-doped $WS_2$ monolayers [52], and ~4 at.% for V-doped $WSe_2$ monolayers [53]). Unlike the case of V-$WS_2$ monolayers (Fig. 4b,c), we have observed the thermally induced spin flipping (TISF) phenomenon in V-$WSe_2$ monolayers (Fig. 4e) due to the presence of antiferromagnetic coupling between spins at V-sites and their nearest W sites (Fig. 4f). Interestingly, the TISF phenomenon can be achieved at low magnetic fields (less than 100 mT) and manipulated by modifying the vanadium concentration within the $WSe_2$ monolayer. These 2D DMSs can thus be used as novel 2D spin filters to enhance the spin to charge conversion efficiency of spin-caloritronic devices [84,109]. It has been reported that after magnetic transition metal (e.g., V or Fe) doping, the photoluminescence signal is strongly suppressed in 2D-TMDs [52,53,55,61], which has been attributed to the formation of impurity energy bands caused by p-type doping at the valence band maximum [51-53,72]. Therefore, it is crucial to select appropriate doping concentrations at which 2D-TMDs exhibit optimized magnetic and optical properties.

Duong *et al.* studied the effect of electric gating on the magnetization of V-doped $WSe_2$ monolayers and found that hole injection enhances the magnetization, while electron injection significantly decreases it [54,57]. Complementing experimental findings, DFT calculations reveal the dominant hole-mediated long-range ferromagnetic interactions between V-spins in atomically thin V-doped TMD systems [57]. Our previous studies have shown that, after vanadium doping, significant photoluminescence is still present in 2 at.% V-doped $WS_2$ [52] and 4 at.% V-doped $WSe_2$ [53] monolayers, which both exhibit the largest saturation magnetization ($M_S$) values among the compositions investigated. These observations have led us to propose that the ferromagnetism in the V-$WS_2$ or V-$WSe_2$ monolayer can be mediated by illumination with a laser of appropriate energy, specifically, above the optical gap (Fig. 3c). Electrons from photogenerated electron-hole pairs may be captured by the V atoms, thus creating an imbalance in the carrier population (i.e., the generation of excess holes) such that the ferromagnetism of the monolayer is modified (Fig. 3d-f).

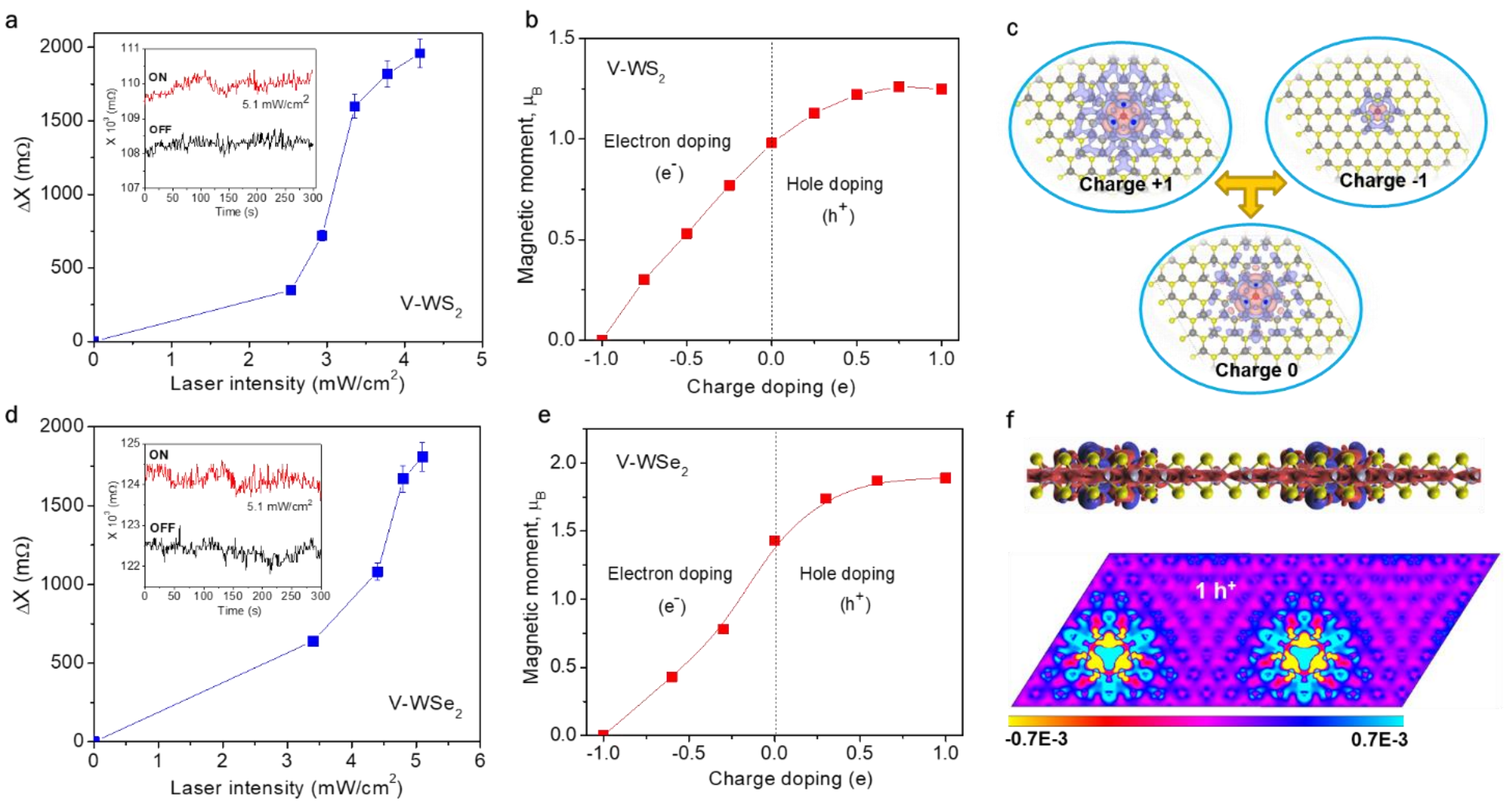

**Figure 5.** The laser intensity dependence of the change in reactance ($\Delta X$) of (**a**) V-$WS_2$ and (**d**) V-$WSe_2$ monolayers. Insets show the change in reactance upon light illumination with a 650 nm laser for (**a**) V-$WS_2$ and (**d**) V-$WSe_2$ monolayers; The net magnetic moments of (**b**) V-$WS_2$ and (**e**) V-$WSe_2$ monolayers with different carrier doping densities; (**c**) The projected magnetic moment along the c-axis of the V-$WS_2$ monolayer upon one hole and one electron injection relative to no charge injection; (**f**) The projected magnetic moment along the c-axis of the V-$WSe_2$ monolayer upon a single hole injection. Hole injection increases the magnetization of the V-$WS_2$ or V-$WSe_2$ monolayer. Panels (a,b,c) taken with permission from Ref. [56]; Panel (f) taken with permission from Ref. [57].

The combination of magnetic and semiconducting properties in 2D-TMD DMSs has indeed enabled light modulation and tunability of magnetization, as demonstrated for V-$WS_2$ and V-$WSe_2$ monolayers (Fig. 5). As can be clearly seen from Fig. 5(a,d), both V-$WS_2$ and V-$WSe_2$ systems exhibit a similar light intensity-dependent magnetization trend. Note that undoped TMD samples (pristine $WS_2$ and $WSe_2$ monolayers) do not exhibit light-mediated magnetism, and the observed enhancement of the magnetization in illuminated V-doped TMD monolayers is not due to a laser/sample heating effect but originates from carrier-mediated ferromagnetism, similar to the case of a p-type (In,Mn)As/GaSb semiconductor [110]. DFT calculations demonstrate that hole injection shifts the Fermi level deeper inside the valence band, while electron injection shifts it towards the conduction band edge [56,57]. As a result, the magnetic moment becomes larger with increasing hole concentration, while an opposite trend is observed for electron injection (Fig. 5b,e). Increasing the concentration of holes results in a more robust magnetic moment across the lattice, where W atoms far from the V site exhibit an enhanced magnetic moment. Since long-range ferromagnetic interactions are mediated by free holes in V-$WS_2$ and V-$WSe_2$ systems it is unsurprising that optically injecting hole-carriers leads to enhanced ferromagnetism. The theoretical calculations fully support the experimental findings. It is worth noting that at large hole concentrations, the magnetic moment saturates, confirming the feature observed experimentally (Fig. 5a,d). This has been attributed to the screening of charge carriers at high hole concentrations. The experimental results presented in Fig. 5 were obtained from MLCR experiments conducted using a 650 nm laser. We also performed MLCR measurements on the same samples using a 520 nm laser and observed enhanced magnetization, confirming that light-enhanced magnetization can be achieved with any wavelength above the optical gap.

Although the V-$WS_2$ and V-$WSe_2$ systems share a similar light-mediated magnetization effect (Fig. 5a,d), a noticeable difference in magnetic coupling between spins at a V-site and its nearest W-sites (namely, the nearest V-W spins) is evident [56,57]. DFT calculations reveal that interactions between the nearest V-W spins are *ferromagnetic* in V-$WS_2$ monolayers [52,56] but *antiferromagnetic* in V-$WSe_2$ monolayers at the same V-doping level [53,57]. In the case of V-$WS_2$ monolayers, the ferromagnetic interaction between the nearest V-W spins becomes stronger with increasing hole concentration (or increase of light intensity), giving rise to an enhanced magnetic moment (Fig. 5c). However, the situation is rather different for V-$WSe_2$ monolayers (Fig. 5f), in which the V atom couples antiferromagnetically to the nearest W sites, and ferromagnetically to the distant W-sites. The introduction of charge carriers mediates this interaction, where increasing hole carriers results in an enhanced magnetic moment at the V site. Additionally, the magnetic moment at the near W sites, flips from antiferromagnetic to weakly ferromagnetic, which combined with the increasing magnetic moment at the V site, results in the enhanced long-range ferromagnetism (Fig. 5f). What is particularly striking in the V-$WSe_2$ system, is the antiferromagnetic coupling between V and near W sites, which appears to be responsible for the lack of saturation of the magnetic moment at higher laser intensities (Fig. 5d).

These findings suggest that the light-mediated magnetism effect is *universal* to the class of 2D-TMD DMS and should be fully exploited for applications in opto-spintronics, opto-spin-caloritronics, spin-valleytronics, and quantum communications.

### *3.2. Two-Dimensional Transition Metal Dichalcogenide Heterostructures*

Typical vdW TMD monolayers offer extensive flexibility and integration with one another [7,8,15,17,51]. Stacking different 2D-TMDs can create novel heterostructures with atomically sharp interfaces and properties that would otherwise be absent in their individual components [15,17,51]. Recent studies have demonstrated that the magnetic or magneto-optical properties of a non-magnetic

TMD (e.g., $WS_2$, $MoS_2$) can be induced or enhanced by stacking it with another magnetic TMD ($VS_2$, $VSe_2$, $CrSe_2$, $MnSe_2$) [13,50,111-113]. This occurs as a result of combined charge transfer and magnetic proximity (PM) effects [13,50,80,113]. By forming $MoS_2/VS_2$ and $WS_2/VS_2$ interfaces, DFT calculations indicate that charge transfer occurs across the interface from $MoS_2$ or $WS_2$ to $VS_2$, as both $MoS_2$ and $WS_2$ have smaller work functions compared to $VS_2$ [113]. Electrons accumulate in the $VS_2$ layer, while holes occupy the $MoS_2$ or $WS_2$ layer. Consequently, both $MoS_2/VS_2$ and $WS_2/VS_2$ heterostructures exhibit enhanced magnetic properties, with Curie temperatures exceeding 300 K [113]. By forming $CrSe_2/WSe_2$ interfaces, DFT calculations by Li *et al.* also show that charge transfer from the $WSe_2$ to the $CrSe_2$ layer and interlayer coupling within $CrSe_2$ play crucial roles in the magnetic properties of the heterostructure [50]. A similar situation is anticipated in $VSe_2/MoS_2$ and $MnSe_2/MoTe_2$ systems [13,112], where both $VSe_2$ and $MnSe_2$ monolayers have been reported to display ferromagnetism above room temperature [13,14]. The combined semiconducting and ferromagnetic properties make these heterostructures appealing for opto-spintronics and opto-spin-caloritronics, as the application of external stimuli such as electric gating and light is likely to promote the charge transfer process and hence alter the magnetic and magneto-optic properties of the heterostructures [87].

As demonstrated above for V-$WS_2$ and V-$WSe_2$ monolayers, the magnetic and semiconducting properties coexist, giving rise to their magneto-optical properties and granting access to the rich electronic properties of 2D semiconductors, enabling the light tunability of magnetization. These findings suggest a similar phenomenon in $VSe_2/WS_2$ and $VSe_2/MoS_2$ heterostructures. An intriguing feature of the $VSe_2/WS_2$ or $VSe_2/MoS_2$ heterostructure is the presence of strong interfacial magnetic coupling and the potential for charge transfer between the two layers, which may play a key role in mediating the magnetization of the film. Bilayer $MoS_2$ shows a strong photon absorption peak around 600-700 nm, and by illuminating the $VSe_2/MoS_2$ film with energy close to the peak, we expect considerable photogeneration of electron-hole pairs. These pairs may be separated through the electric field at the heterointerface, which leads to the observed light-tunable magnetism in the film.

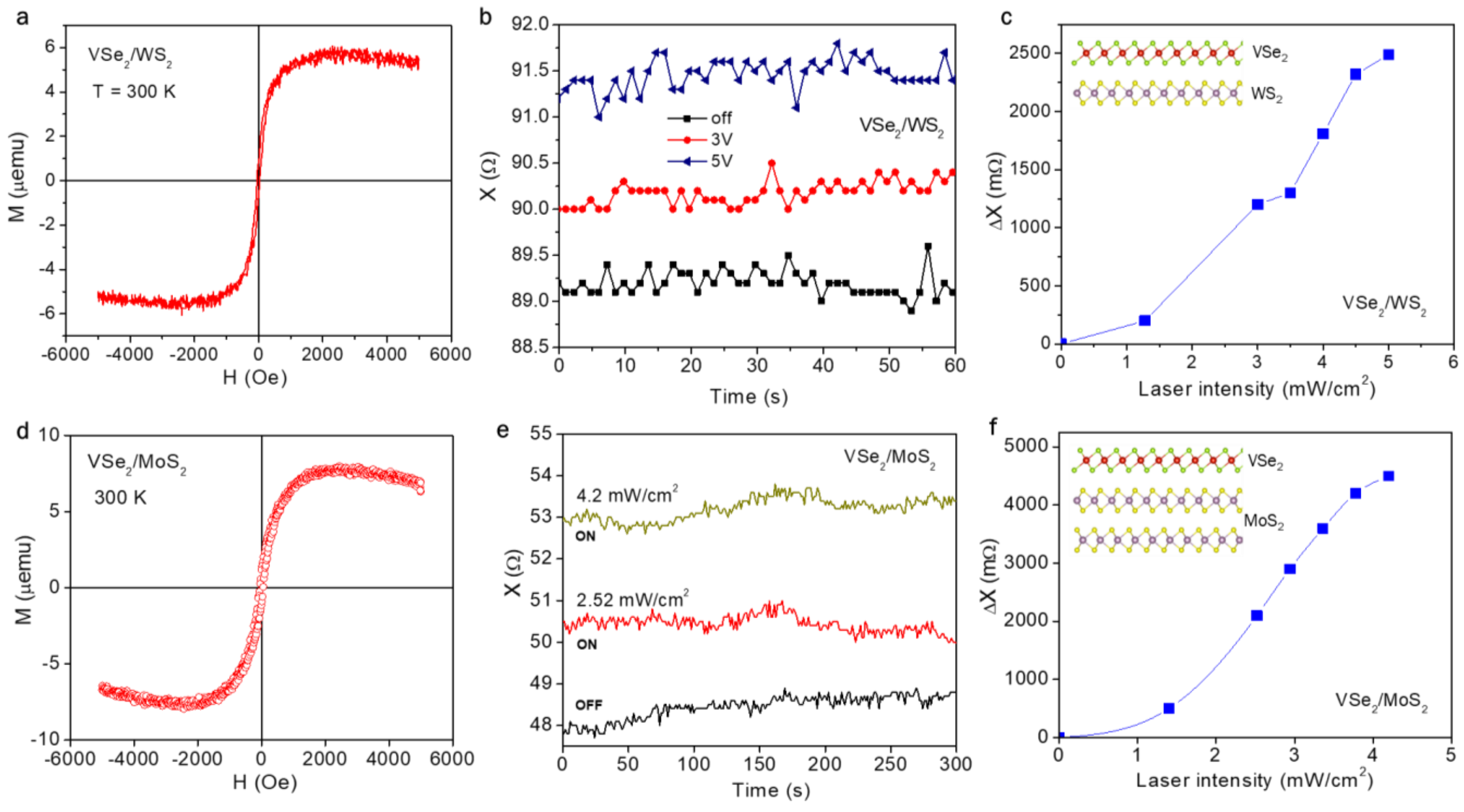

**Figure 6.** Magnetic hysteresis loops (*M*(*H*)) taken at 300 K for (**a**) 1L-$VSe_2$/1L-$WS_2$ and (**d**) 1L-$VSe_2$/2L-$MoS_2$ films. The reactance (*X*) vs. time upon illumination with a 650-nm laser with various intensities for (**b**) 1L-$VSe_2$/1L-$WS_2$ and (**e**) 1L-$VSe_2$/2L-$MoS_2$ films. Laser intensity dependence of the reactance change ($\Delta X$) for (**c**) 1L-$VSe_2$/1L-$WS_2$ and (**f**) 1L-$VSe_2$/2L-$MoS_2$ films.

To investigate this hypothesis, we conducted MLCR experiments on both $VSe_2$/$WS_2$ and $VSe_2$/$MoS_2$ heterostructure samples upon light illumination using a diode laser with a wavelength of ~650 nm ($h\nu$ ~ 1.91 eV). In this case, the $VSe_2$/$WS_2$ or $VSe_2$/$MoS_2$ heterostructure consists of a vertically stacked monolayer (1L) $VSe_2$ and monolayer (1L) $WS_2$ or bilayer (2L) $MoS_2$ grown on an $SiO_2$ substrate by combining molecular beam epitaxy (MBE) and chemical vapor deposition (CVD), respectively. Representative results of the $VSe_2$/$WS_2$ and $VSe_2$/$MoS_2$ samples are displayed in Fig. 6. It is noteworthy to observe that both the 1L-$VSe_2$/1L-$WS_2$ and 1L-$VSe_2$/2L-$MoS_2$ samples exhibit a pronounced ferromagnetic signal at room temperature (Fig. 6a,d), as well as light-tunable ferromagnetism at room temperature (Fig. 6b,e). For both heterostructures, the magnetization significantly increases with increasing laser intensity and tends to saturate at high laser intensities (Fig. 6c,f). The light intensity dependence of magnetization for the $VSe_2/WS_2$ and $VSe_2/MoS_2$ heterostructures (Fig. 6c,f) is similar to that observed for V-doped TMD monolayers (Fig. 5a,d) and for a $CH_3NH_3PbI_3$/LSMO heterostructure [96]. The enhancement and tunability of light-mediated ferromagnetism in the 1L-$VSe_2$/2L-$MoS_2$ film were also independently confirmed by the pump-probe MOKE technique [90]. It is important to note that this light-mediated magnetism effect is not observed in the individual layers ($WS_2$, $MoS_2$, $VSe_2$); only when they are stacked together do we observe light-dependent magnetization. These results demonstrate the universality of the light-mediated magnetism effect and pave a new pathway for the design and fabrication of novel van der Waals heterostructures for use in 2D van der Waals spintronics, opto-spin-caloritronics, and opto-valleytronics.

Comparison of the light-mediated magnetization results of 1L-$VSe_2$/2L/$MoS_2$ and 1L-$VSe_2$/BSC $MoS_2$ (BSC: Bulk single crystal) samples has shown that the change in magnetization in 1L-$VSe_2$/2L-$MoS_2$ is approximately 4 times greater than that of 1L-$VSe_2$/BSC-$MoS_2$ (Fig. 7a) and this result is reproducible (inset of Fig. 7a). This observation leads us to believe that electron confinement effects on the 2L-$MoS_2$ might play a significant role in the mechanism behind light mediated magnetism in this heterostructure (Fig. 7b), as compared to the case of BSC $MoS_2$ (Fig. 7c).

To elucidate the mechanism of light-enhanced magnetism in the 1L-$VSe_2$/2L-$MoS_2$ system, we conducted first-principles defect calculations [114] based on the Heyd-Scuseria-Ernzerhof (HSE) hybrid functional (with the mixing parameter set to 0.15) [115] as implemented in VASP [116] and with the van de Waals correction [117] and finite-supercell size effects [118] included. The calculations suggest that this enhancement can be tentatively attributed to the presence of sulfur vacancies ($V_S$) in 2L-$MoS_2$ (Fig. 7d-f). $V_S$ are found to be stable in their negative charge states ($V_S^-$) within the range of Fermi-level values closer to the conduction-band minimum (CBM). Under n-type conditions, $V_S^-$ represents the lowest-energy magnetic native point defect ($1\mu_B$ per $V_S^-$) (Fig. 7e). The magnetic interaction between the two $V_S^-$ defects is weakly ferromagnetic. In the $VSe_2/MoS_2$ system, the presence of the $Vse_2$ layer with a larger work function of ~4.5 eV (compared to ~4.1 eV for the $MoS_2$ layer) results in an accumulation of electrons in the $Vse_2$ layer, creating a depleted region in the $MoS_2$ side of the hetero-interface and subsequent formation of a Schottky barrier. The mechanism for ferromagnetism in the $Vse_2/MoS_2$ system remains under debate; however, the accumulation of electrons in the $Vse_2$ layer could give rise to enhanced ferromagnetism in this layer and thus contribute to the change in net magnetization of the $Vse_2/MoS_2$ heterostructure when the material is not exposed to light. It should be noted that 1L-$Vse_2$ has a more dominant magnetic contribution to the net magnetization of the 1L-$Vse_2$/2L-$MoS_2$ film.

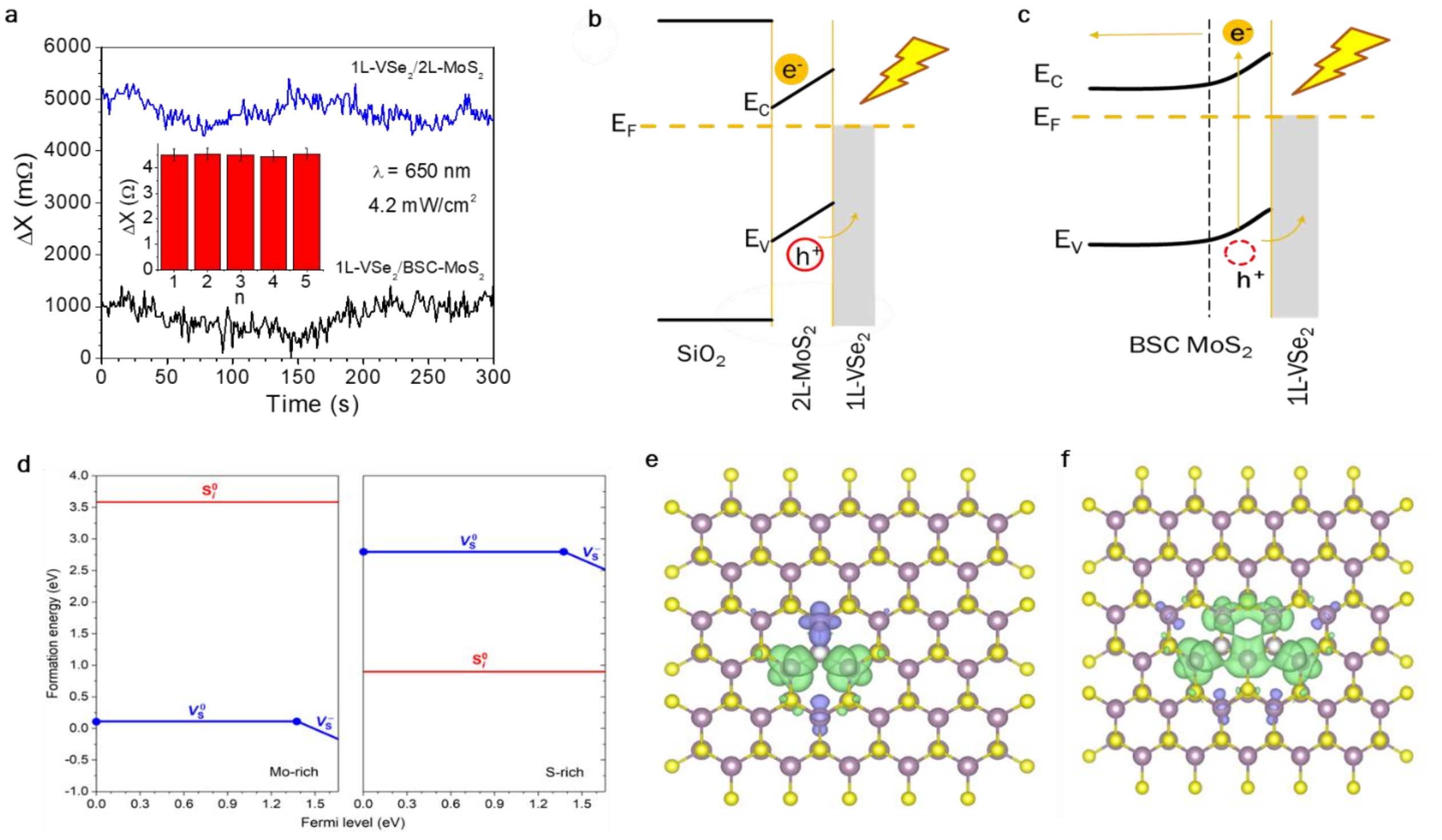

**Figure 7.** (**a**) Comparison of the reactance change ($\Delta X$) due to light irradiation at the same wavelength ($\lambda$ = 650 nm) and intensity (4.2 mW/cm$^2$) between the 1L-$Vse_2$/2L-$MoS_2$ film and the 1L-$Vse_2$/BSC-

$MoS_2$ film. Upon light irradiation, the enhancement of the magnetization is approximately 4 times greater in the 1L-$Vse_2$/2L-$MoS_2$ film than in the 1L-$Vse_2$/BSC-$MoS_2$ film. **Inset of (a)** shows values of the reactance change ($\Delta X$) measured at different times demonstrating the reproducibility of the observed effect. Schematics show a possible charge transfer via (**b**) the 2L-$MoS_2$/$Vse_2$ and (**c**) BSC-$MoS_2$/$Vse_2$ interface upon light irradiation; (**d**) Formation energies of sulfur vacancies ($V_S$) and interstitials ($S_i$) in a 2*H*-$MoS_2$ bilayer under extreme Mo-rich and S-rich conditions and plotted as a function of the Fermi level from the VBM to the CBM. The slope of energy-line segments indicates the charge state. The vacancy (interstitial) is found to be the dominant native point defect under the Mo-rich (S-rich) condition. $V_S$ is stable as the neutral (and nonmagnetic) vacancy $V_S^0$ in a wide range of Fermi-level values and as the negatively charged vacancy $V_S^-$ near the CBM (i.e., under n-type conditions). The (0/–) transition level of $V_S$ is at 0.31 eV below the CBM. $V_S^-$ has a calculated magnetic moment of $1\mu_B$. $S_i$ is, on the other hand, stable as the neutral and nonmagnetic vacancy $S_i^0$ in the entire range of Fermi-level values; (**e**) Top-view of spin density associated with a single negatively charged defect $V_S^-$; (**f**) top-view of the spin density associated with a pair of $V_S^-$. Large (purple) spheres are Mo, and small (yellow) spheres are S. The lattice site of the vacancy is marked by a small (white) sphere. The isosurface level is set to 0.014 e/Å$^3$ and the green (purple) isosurfaces correspond to up (down) spin.

Upon light illumination, electron-hole pairs are generated and subsequently separated by the interfacial electric field at the Schottky barrier between the two materials. Consequently, an accumulation of excited electrons builds up in the conduction band of $MoS_2$. This 2D electron gas can increase the negative charge of the sulfur vacancies in the $MoS_2$ layer, which, in turn, increases the net magnetic moment (Fig. 7f). For instance, the nonmagnetic $V_S^0$ becomes $V_S^-$, for a certain duration (during laser irradiation). Our calculations indicate that a portion of the extra electrons will localize on the vacancy. At higher laser powers, more photo-generated electrons in the conduction band of $MoS_2$ will completely fill the available confined-states and eventually leak to the $VSe_2$ layer, which could explain the saturation of magnetization enhancement. Due to the two-dimensional nature of 2L-$MoS_2$, both photo-generated electrons and sulfur vacancies are confined to the vicinity of the heterostructure interface. This facilitates the ferromagnetic interaction between nearby neighboring vacancies, which not only leads to a larger magnetic moment in the film but also enhances light-controlled magnetism. This suggests that besides the interfacial coupling of the heterostructure, the electron confinement and the concentration of sulfur vacancies also mediate the change in magnetization of the film under illumination (see, Fig. 7a-c). This effect is significantly reduced when the thickness of the $MoS_2$ layer is increased, as demonstrated by the change in magnetization of the BSC-$MoS_2$/$VSe_2$ being 4 times smaller than that of 2L-$MoS_2$/$VSe_2$ (Fig. 7a). We attribute this to the lack of confinement near the $MoS_2$/$VSe_2$ interface. In the case of BSC-$MoS_2$/$VSe_2$ (Fig. 7c), photogenerated electrons are no longer confined to the interface but can move deeper into the $MoS_2$ layer, significantly reducing the concentration of electrons near the interface. Nevertheless, further studies are required to fully understand the mechanism of light-mediated ferromagnetism in the $VSe_2$/$MoS_2$ system. It should also be noted that sulfur vacancies in 2L-$MoS_2$ are likely not the sole source of magnetism; other point defects and extended defects in the samples may play a role. Our theoretical argument regarding light-enhanced ferromagnetism is applicable to any defect whose electrical and magnetic properties resemble those of sulfur vacancies.

## 4. Opportunities and Challenges

Our findings have established that the light-modulated magnetism effect is *universal* to 2D-TMD DMSs, including V-doped TMD monolayers (V-$WS_2$, V-$WSe_2$, and V-$MoS_2$). In addition to their electrically tunable magnetic functionalities, the optically tunable atomic-level magnetism at room temperature makes 2D-TMD DMSs even more attractive for applications in spintronics, spin-caloritronics, and valleytronics. It is worth noting that both electrons and holes are populated in light-illuminated MLCR experiments, while the DFT simulations consider the separated effect for each type of carrier. The coexistence of magnetic and semiconducting properties is necessary; therefore, higher doping concentrations are unlikely to exhibit this effect, as their semiconducting qualities are strongly suppressed. On the other hand, lower V-doping concentrations are interesting to study, since excellent semiconducting properties are preserved while the magnetic moment is slightly smaller compared to optimally V-doped TMD samples. Investigating the influence of different V doping concentrations on 2D-TMDs' light-mediated magnetism effect is an intriguing direction for future research. Further studies are necessary to understand the relationship between doping concentration and light-enhanced magnetism.

In addition, we have demonstrated the universality of the light-mediated magnetism effect in 2D-TMD ferromagnet/semiconductor heterostructures, including $VSe_2/WS_2$ and $VSe_2/MoS_2$ systems. Charge transfer, proximity, and confinement effects play a crucial role in enhancing light-mediated magnetization in these 2D systems. However, other effects such as interdiffusion, intercalation, twisting, and moiré patterns, which could occur during heterostructure formation remain largely unexplored [31,45,51,78,99,119]. Recently, Wang *et al.* revealed the novel possibility of tuning spin-spin interactions between moiré-trapped holes using optical means, inducing a ferromagnetic order in $WS_2/WSe_2$ moiré superlattices [99]. In the case of $VSe_2/MoS_2$ heterostructures, the moiré pattern has been observed at low temperatures (below the charge density wave transition temperature, $T_{CDW}$), and the presence of this moiré pattern led to enhanced magnetization, strong interfacial magnetic coupling, and the exchange bias (EB) effect within this temperature range [13,85]. To fully understand the moiré effect, and to exploit the light tunability of the EB effect, it would be interesting to investigate the light-mediated magnetism effect in $VSe_2/MoS_2$ and $VSe_2/WS_2$ heterostructures at $T < T_{CDW}$. DFT calculations suggest that charge transfer from the ferromagnetic metal $VSe_2$ ($CrSe_2$) to the semiconductor $MoS_2$ ($WSe_2$) gives rise to the magnetic moment of the $MoS_2$ ($WSe_2$) layer. A fundamental question emerges: *Will a similar effect occur when the 2D semiconducting TMD ($MoS_2$, $WS_2$, etc.) layer interfaces with a non-magnetic metal like graphene?* Exploring magnetism and light effects in semiconducting 2D-TMDs (both pristine and magnetically doped TMDs) interfaced with graphene will not only address this important question but also provide new insights into charge transfer-mediated magnetism in light-illuminated van der Waals heterostructures composed of 2D-TMD DMSs and other 2D materials. Depending on the work function difference between the two component materials, holes/electrons can be transferred into 2D-TMD DMSs, resulting in enhanced or reduced magnetization. This represents a promising, innovative approach for designing novel 2D-TMD heterostructures with enhanced magnetic and magneto-optic properties through a combined chemical doping and interface engineering (via the light-modulated directional charge transfer mechanism) strategy.

From an application perspective, the optically tunable magnetic properties of 2D-TMD DMSs and heterostructures are desirable for opto-spintronics, opto-spin-caloritronics, and valleytronics. Ghiasi *et al.* demonstrated charge-to-spin conversion across a monolayer $WS_2$/graphene interface due to the Rashba-Edelstein effect (REE) [120]. Alternatively, using 2D-TMD DMSs such as V-$WS_2$ and V-$WSe_2$ monolayers may not only boost spin-charge conversion efficiency but also enable optical manipulation of the spin-charge conversion process. A comprehensive review by Sierra *et al.* highlighted the novel application perspectives of opto-spintronics [7].

Thermally induced spin currents based on the spin Seebeck effect (SSE), a phenomenon discovered by Uchida *et al.* [121,122], laid the foundation for a new generation of spin-caloritronic devices. A pure spin current can be generated in a ferromagnetic (FM) material (like YIG: $Y_3Fe_5O_{12}$) due to a built-up electric potential across a temperature gradient upon the application of a magnetic field. This spin current can be converted into a technologically useful voltage via the inverse spin

Hall effect of a heavy metal (HM) with strong spin-orbit coupling (like Pt) in an FM/HM structure. Recently, Kalappattil *et al.* showed that inserting a thin (~5 nm) organic *semiconducting* layer of $C_{60}$ can significantly reduce the conductivity mismatch between YIG and Pt and the surface perpendicular magnetic anisotropy of YIG, resulting in a giant enhancement (600%) in the longitudinal SSE [91]. Following this approach, Lee *et al.* inserted a *semiconducting* $WSe_2$ monolayer between Pt and YIG and observed the giant SSE in Pt/$WSe_2$/YIG [123]. DFT calculations indicate that inserting a 2D-TMD DMS (e.g., V-$WSe_2$) in an FM/HM bilayer system not only reduces the conductivity mismatch but also enhances the spin mixing conductance and hence the spin-to-charge conversion efficiency via the SSE [123]. By taking advantage of the light-tunable magnetization of 2D-TMD DMSs [56,90], Phan *et al.* recently proposed a new route for the optical control of thermally induced spin currents through 2D-TMD DMS interfaces in FM/HM systems, establishing the new subfield named "Opto-spin-caloritronics" [84], which can harness "*light as the new heat*". Further studies are needed to fully exploit this potential.

Semiconducting 2D-TMDs (e.g., $WSe_2$, $WS_2$, $MoS_2$) are excellent candidates for use in valleytronic devices [48,49]. Controlling and manipulating the valley polarization states in these 2D-TMDs using external stimuli (optic, electric, and magnetic field) is essential [77,124-127]. Doping magnetic atoms (Fe, V) into a $MoS_2$ monolayer to form Fe-$MoS_2$ [66] or V-$MoS_2$ [67] DMSs has been reported to enhance valley splitting in $MoS_2$ monolayers. Since 2D-TMD DMSs exhibit strong magnetic responses to both electric fields and lasers [56,57], their valleytronic properties can be manipulated by these external stimuli. Seyler *et al.* [41] experimentally exploited light to control $CrI_3$ magnetization in a $CrI_3$/$WSe_2$ heterostructure, demonstrating the optical modulation of valley polarization and valley Zeeman splitting within the $WSe_2$ monolayer. However, the $CrI_3$ monolayer exhibits ferromagnetic ordering below 50 K, rendering the $CrI_3$/$WSe_2$ heterostructure impractical for use in valleytronic devices that operate at ambient temperatures. In this context, the optical modulation of magnetic and valleytronic properties of $VSe_2$($MnSe_2$)/$MoS_2$ and $VSe_2$($MnSe_2$)/$WS_2$ heterostructures appears more compelling, as the $VSe_2$ or $MnSe_2$ layer exhibits ferromagnetic order above room temperature [13,14]. Based on DFT calculations, He *et al.* showed that an ultrafast laser pulse can induce a ferromagnetic state in a nonmagnetic $MoSe_2$ monolayer when interfaced with the $MnSe_2$ monolayer, which orders ferromagnetically above room temperature [128]. Such ultrafast optical control of 2D magnetism is highly compelling for applications in ultrafast spintronics and magnetic storage information technology [129]. It is worth noting that the magnetic properties of 2D-TMDs have contributions from defect- and dopant-induced magnetic moments and their couplings [51,60]. Therefore, it would be of significant interest to investigate the effects of transition metal or chalcogen vacancies and magnetic dopant concentrations on the magnetic, magneto-optic, and valleytronic properties of free-standing TMD monolayers, as well as those placed on magnetic substrates in heterostructures. Scheme 1 highlights opportunities and challenges in exploiting 2D-TMD DMSs and heterostructures for use in modern nanodevices.

**Scheme 1.** Opportunities and challenges in the research of 2D-TMD magnets.

<table>
<tr><th colspan="2">Two-dimensional van der Waals Dilute Magnetic Semiconductors</th></tr>
<tr><th>Opportunities</th><th>Challenges</th></tr>
<tr><td>

- Electrically and optically tunable magnetism in 2D-TMD DMSs (e.g., V-doped $WSe_2$) for spin transistors, sensors, and spin-logic device applications

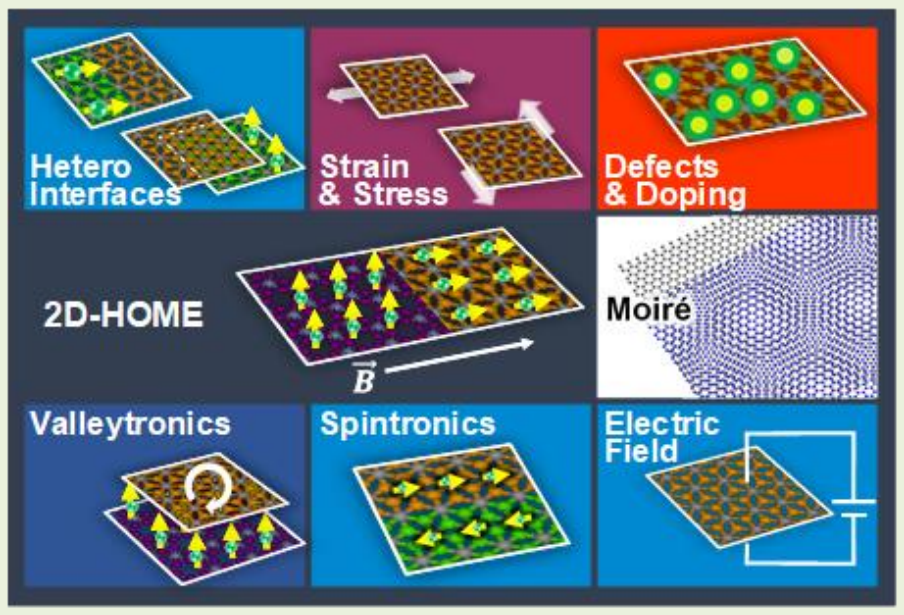

- Optically tunable twisted and moiré magnetism and related phenomena in 2D-TMD heterostructures (e.g., $VSe_2/MoS_2$, $WS_2/WSe_2$) for spintronics and opto-spintronics
- Exploring light-tunable exchange bias or exchange anisotropy in 2D vdW heterostructures (e.g., $VSe_2/MoS_2$, $VS_2/WS_2$) for opto-spintronics
- Optically tunable magnetism and spin-thermo-transport (SSE, ANE, etc.) in

</td><td>

- Weak ferromagnetism
- Difficulty controlling magnetic dopants/defects in 2D-TMDs to achieve reproducible magnetic properties (e.g., $M_S$, $H_C$)
- Strong suppression of ferromagnetism and photoluminescence in heavily doped 2D-TMDs
- Air instability (magnetic signals are degradable when exposed to air)
- Difficulty with integration and fabrication of 2D-TMD-based devices
- Air instability (magnetic degradation causes the SSE voltage to reduce)
- Gaps in understanding spin-charge-phonon coupling mechanisms that govern spin transport across 2D vdW interfaces
- Large discrepancy between the theoretical and experimental values of SSE voltage for the reported 2D magnets and heterostructures

</td></tr>
</table>

<table>
<tr>
<td>
FM/2D-TMD/HM structures (e.g., YIG/ML V-doped $WSe_2$/Pt) for opto-spin-caloritronics<br>
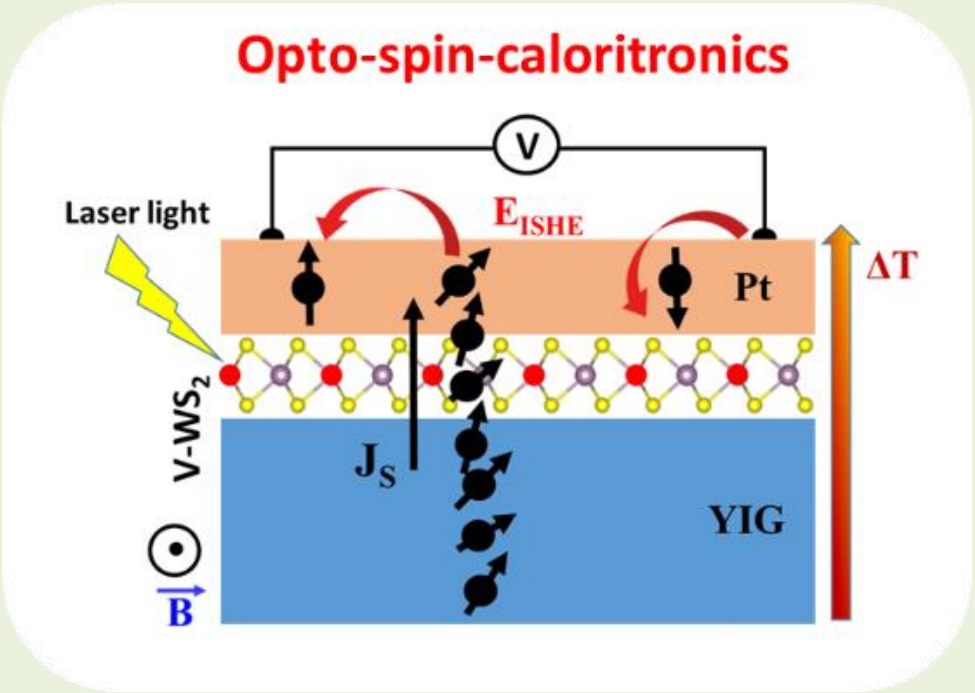

<ul><li>Ultrafast magnetism and ultrafast spin-thermo-transport for ultrafast opto-spin-caloritronic device applications</li></ul>
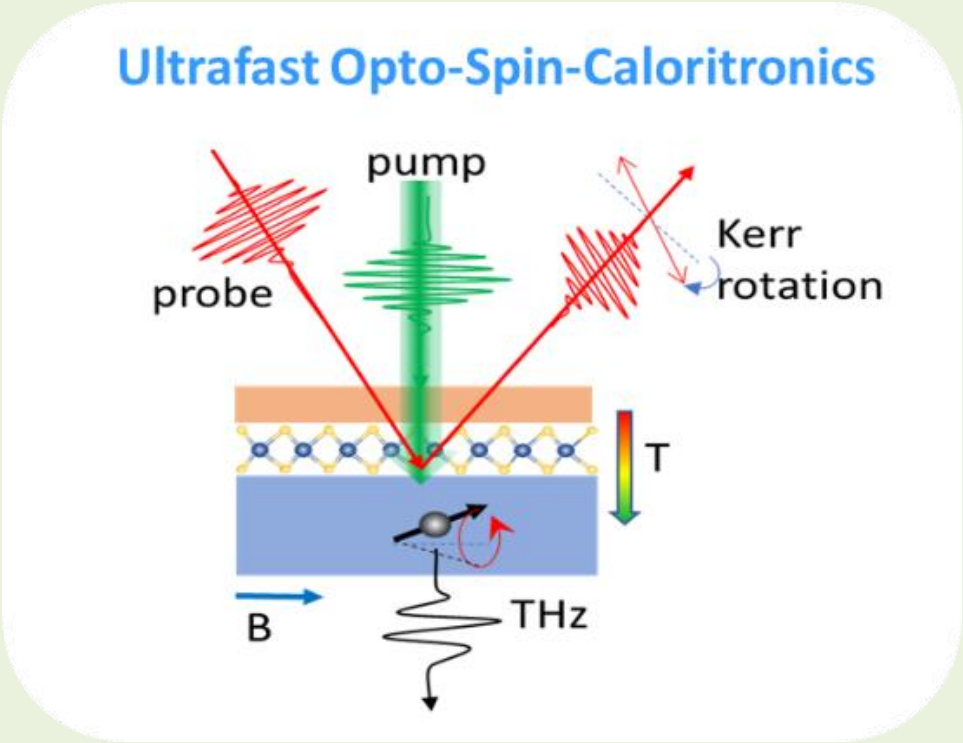

</td>
<td>
<ul>
<li>Need to select vdW materials that order magnetically at high temperatures ($T_C$, $T_N$) to achieve EB effects around room temperature</li>
<li>Difficulty synthesizing and stacking vdW magnetic materials into heterostructures</li>
<li>Surface and interface oxidization (magnetic degradation)</li>
<li>Undesired effects of intercalation, interdiffusion at vdW magnet/metal interfaces, twisting, moiré patterns, etc.</li>
<li>MOKE signals depend on quality of surfaces and interfaces of materials used; nonlocal heating effects on spin dynamics and transport</li>
<li>Complex device structures and short lifetime of the devices</li>
</ul>
</td>
</tr>
</table>

## 5. Concluding Remarks and Outlook

We have established the universality of the light-modulated magnetization effect in 2D-TMD DMSs, including V-doped TMD monolayers (V-$WS_2$, V-$WSe_2$, V-$MoS_2$). This effect is attributed to the presence of excess holes in the conduction and valence bands, as well as carriers trapped in the magnetic doping states, which mediate the magnetization of the TMD layer. Additionally, we have demonstrated the universality of the light-mediated magnetism effect in 2D-TMD ferromagnet/semiconductor heterostructures such as $VSe_2/WS_2$ and $VSe_2/MoS_2$. This effect is attributed to photon absorption at the TMD layer (e.g., $WS_2$, $WSe_2$, $MoS_2$), generating electron-hole pairs that mediate the magnetization of the heterostructure. These findings pave a new pathway for the design of novel 2D-TMD van der Waals heterostructures that exhibit unique magneto-optical coupling functionalities that enable the next generation of high-performance optoelectronics, ultrafast opto-spintronics, opto-spin-caloritronics, valleytronics, and quantum technologies.

We have demonstrated the importance of proximity, charge-transfer, and confinement effects in enhancing light-mediated magnetization in 2D-TMD heterostructures, but other effects such as interdiffusion, intercalation, twisting, and moiré patterns may also be significant [31,45,51,78,99,119]. Further studies are thus needed to fully understand these effects. It appears that 2D-TMD magnetism arises from multiple contributions of vacancy- and dopant-induced magnetic moments, as well as their magnetic couplings, whose strengths vary depending on their complex vacancy-dopant configurations [51,60]. Understanding how light mediates magnetization in 2D-TMDs with controlled dopant/vacancy concentrations is critical. The "twisting" effect has been reported to significantly alter the magnetic and valleytronic properties of 2D-TMDs [130-135]. Twisting graphene from a 2D-TMD in a 2D-TMD/graphene heterostructure can enhance the valley Zeeman and Rashba effects [130,131], as well as the charge-to-spin conversion efficiency [132,133]. By tailoring the atomic interface between twisted bilayer graphene and $WSe_2$, Lin *et al.* showed strong electron correlation within the moiré flat band, which stabilizes insulating states at both quarter and half filling, and the spin-orbit coupling drives the Mott-like insulator into ferromagnetism [134]. In addition to the magnetic proximity and charge transfer effects, twisting adds an interesting experimental knob to tune the magnetic and magneto-optic functionalities of 2D-TMDs for spintronics and valleytronics applications.

From an application standpoint, the room-temperature electrically and optically tunable magnetic properties make 2D-TMD DMSs excellent candidates for use in spin transistors, logic, and magnetic memory devices [51,54,86]. In opto-spintronics, ultrafast optical control of 2D magnetization may yield the fastest information recording and processing with minimal dissipative power [7,129]. Experimental studies are needed to verify these theoretical predictions [73,113]. Additionally, 2D-TMD DMSs can serve as novel 2D spin filters to boost the spin-to-charge conversion efficiency via the SSE in FM/2D-DMS/HM systems [84,109]. A comprehensive understanding of the spin-charge-phonon coupling mechanisms in such 2D spin filters is currently lacking but crucial for unlocking the potential of "Opto-Spin-Caloritronics," which warrants further study.

Compared to their 2D-TMD counterparts, 2D-TMD DMSs appear more promising for use in valleytronic devices [66,67]. To further enhance the valley splitting in these 2D DMSs, it is possible to interface them with other 2D materials such as graphene. Combining chemical doping and interface engineering (charge transfer and/or strain) approaches can create 2D-TMD DMS/graphene or 2D-TMD DMS/2D-TMD heterostructures with enhanced magnetic, magneto-optic, and valleytronic properties that can be tuned by external stimuli (electric gating, light, strain). All these exciting possibilities will facilitate further research.

To provide insightful guidance on the development of 2D-TMD-based devices, we present in Table 1 a list of promising 2D-TMD magnets and heterostructures. While most of the 2D-TMD DMSs are synthesized using chemical vapor deposition (CVD), some of their heterostructures are grown by molecular beam epitaxy (MBE) or a combination of both techniques [76]. CVD typically produces 2D films with uniformity, low porosity, high purity, and stability, but generates toxic gases during the reaction. MBE enables in-situ preparation of atomically clean substrates with specific surface reconstructions, facilitating the growth of highly epitaxial 2D films, but it can result in more defects during film growth [74-76]. It is essential to advance these techniques for growing defect-free or defect-controllable 2D-TMD DMSs and heterostructures.

As noted earlier, the high concentrations of magnetic dopants or the presence of abundant defects (transition metal/chalcogen vacancies) in 2D-TMD semiconductors lead to the strong suppression of photoluminescence [51-53]. However, co-doping with different metals such as, Co and Cr, in a $MoS_2$ monolayer has been reported to enhance both photoluminescence intensity and saturation magnetization [71]. Combining chemical co-doping and interface engineering represents a promising strategy for the design of 2D-TMD DMSs with enhanced magnetic and magneto-optic properties for spintronics, opto-spintronics, opto-spin-caloritronics, valleytronics, and quantum communications.

## Acknowledgements

Research at USF was supported by the U.S. Department of Energy, Office of Basic Energy Sciences, Division of Materials Sciences and Engineering under Award No. DE-FG02-07ER46438 and the VISCOSTONE USA under Award No. 1253113200. M.T. acknowledges support from the Air Force Office of Scientific Research (AFOSR) through grant No. FA9550-18-1-0072 and the NSF-IUCRC Center for Atomically Thin Multifunctional Coatings (ATOMIC). This work used computing resources of the Center for Computationally Assisted Science and Technology (CCAST) at North Dakota State University, which were made possible in part by National Science Foundation Major Research Instrumentation (MRI) Award No. 2019077. The authors acknowledge Dr. M. Bonilla and Dr. S. Kolekar for assisting with the synthesis of $VSe_2/MoS_2$ samples.

**Table 1.** 2D-TMD magnets and heterostructures with optically tunable magnetic functionalities for spintronics, spin-caloritronics, and valleytronics.

| Materials | Ordering temperature $T_C$ (K) | Remarks | Ref. |
|---|---|---|---|
| **Metals** | | | |
| $VSe_2$ (1L, 2L) | ~270 – 350 K | Strong magnetism; Sensitive to defects; Less air stability | [13] |
| $MnSe_2$ (1L, 2L) | ~266 – 350 K | Intrinsic magnetism; Less sensitive to defects; Less air stability | [14] |
| $CrSe_2$ (1L) | 50~300 K | Intrinsic magnetism; Less sensitive to defects; Air stability | [50,136] |
| $FeSe_2$ (1L) | ~300 K | Intrinsic magnetism; Air stability | [136] |
| **Semiconductors** | | | |
| V-$WS_2$ (1L) optimal ~2 at.% | ~300 – 400 K | Unexplored optical control of valleytronic states; Less sensitive to air. | [52] |
| V-$WSe_2$ (1L) optimal ~4 at.% | ~300 – 400 K | Unexplored optical control of valleytronic states; Less sensitive to air. | [53] |
| V-$MoSe_2$ (1L) optimal ~2 at.% | ~300 – 400 K | Unexplored optical control of magnetism and valleytronic states; Less sensitive to air. | [63] |
| V-$MoS_2$ (1L) | ~300 – 400 K | Unexplored optical control of magnetism and valleytronic states; Less sensitive to air. | [61,67] |
| V-$MoTe_2$ (1L) | ~300 – 400 K | Unexplored optical control of magnetism and valleytronic states; Less sensitive to air. | [70] |
| Fe-$MoS_2$ (1L) | ~300 – 400 K | Unexplored optical control of magnetism and valleytronic states; Less sensitive to air. | [55,66] |
| Co-$MoS_2$ (1L) | ~300 K | Unexplored optical control of magnetism and valleytronic states; Less sensitive to air. | [58] |
| (Co,Cr)-$MoS_2$ (1L) | ~300 K | Unexplored optical control of magnetism and valleytronic states; Air stability. | [71] |
| Co-$SnS_2$ (SC) | ~120 K | Unexplored optical control of magnetism and valleytronic states; Air stability. | [138] |
| Fe-$SnS_2$ (1L) | ~30 K | Unexplored optical control of magnetism and valleytronic states; Air stability. | [139] |
| Mn-$SnS_2$ (SC) | ~150 K | Unexplored optical control of magnetism and valleytronic states; Air stability. | [140] |
| **Heterostructures** | | | |
| $CrSe_2$/$WSe_2$ (1L/1L) | ~50 – 120 K | Unexplored optical control of magnetism and valleytronic states; Charge transfer; Less sensitive to air. | [50] |
| $VSe_2$/$MoTe_2$ (1L/1L) | ~300 – 350 K | Unexplored optical control of magnetism and valleytronic states; Charge transfer; Less sensitive to air. | [111] |
| $VSe_2$/$MoS_2$ (1L/2L) | ~300 – 350 K | Unexplored optical control of valleytronic states; Observed EB effect; Charge transfer; Less sensitive to air. | [13,85] |
| $MnSe_2$/$MoSe_2$ (1L/1L) | ~300 – 350 K | Unexplored optical control of magnetism and valleytronic states; Charge transfer; Less sensitive to air. | [112] |
| $VS_2/MoS_2$ (1L/1L) | ~300 – 350 K | Unexplored optical control of magnetism and valleytronic states; Charge transfer; Less sensitive to air. | [137] |